# Diamond-based sensing of stray fields from the bulk of thin-film magnets via nano-indentation

*Ming-Zhong Ai*[1,4,5,6,+], *Kang-Yuan Liu*[1,4,5,6,+], *Biao Zhang*[2,+], *Weng-Hang Leong*[1,7], *Yao Gao*[1], *Yue Cui*[1,8], *Guoli Zhu*[1], *Licong Peng*[2], *Yanglong Hou*[2,3,*], *Quan Li*[1,4,5,*], *Ren-Bao Liu*[1,4,5,6,*]

[+]These authors contributed equally: Ming-Zhong Ai, Kang-Yuan Liu, Biao Zhang

[*]Correspondence and requests for materials should be addressed to Y.H. (email: hou@pku.edu.cn), Q.L. (email: liquan@cuhk.edu.hk) or R.-B.L. (email: rbliu@cuhk.edu.hk)

1. Department of Physics, The Chinese University of Hong Kong, Shatin, New Territories, Hong Kong, China
2. Beijing Key Laboratory for Magnetoelectric Materials and Devices, School of Materials Science and Engineering, Peking University, Beijing 100871, China
3. School of Materials, Shenzhen Campus of Sun Yat-Sen University, Shenzhen 518107, China
4. Centre for Quantum Coherence, The Chinese University of Hong Kong, Shatin, New Territories, Hong Kong, China
5. The State Key Laboratory of Quantum Information Technologies and Materials, The Chinese University of Hong Kong, Shatin, New Territories, Hong Kong, China
6. New Cornerstone Science Laboratory, The Chinese University of Hong Kong, Shatin, New Territories, Hong Kong, China
7. Department of Engineering Science, Faculty of Innovation Engineering, Macau University of Science and Technology, Taipa, Macao 999078, China
8. Quantum Science Center of Guangdong-Hong Kong-Macao Greater Bay Area (Guangdong), Shenzhen 518045, China

**ABSTRACT**

Measurement of the magnetization in the bulk of thin-film or two-dimensional materials is important for understanding their intrinsic properties without the complications from edges or domain walls. However, the stray fields from the bulk vanish or are very weak, which limits the application of direct measurement methods. Here, we develop a non-destructive approach to directly measuring the stray fields from the bulk of thin-film magnets at arbitrarily designatable locations, with nanoscale spatial resolution. We employ nano-indentation to induce the leakage of stray fields from the materials and use nano-diamond magnetometers to measure them. We apply the method to iron thin films and determine the intrinsic magnetization in the bulk of the materials. This work provides direct access to the intrinsic magnetic properties of thin-film and low-dimensional materials, as well as a method to study the mechanical effects on magnetization in nanomaterials.

**INTRODUCTION**

Thin-film or two-dimensional magnets are important platforms for studying condensed matter physics [1-3]and have potential applications in quantum spintronics [4-6]. To understand the physics and facilitate technological applications of such materials, it is highly desirable to study the intrinsic magnetization in the bulk of the materials without complications due to edges and/or domain walls. This task, however, is challenging. The thinness and smallness of the samples mean many bulk measurement methods (such as neutron scattering, NMR, X-ray magnetic circular dichroism, and vibrating sample magnetometry) are insufficient [7-10]. The optical Kerr or Faraday rotation provides a non-invasive and direct measurement of magnetization in the bulk of materials but is insensitive to in-plane magnetization and has spatial resolution limited by optical wavelengths [11, 12]. Superconducting quantum interference devices (SQUIDs) have very high sensitivity but low spatial resolution [13, 14]. The magnetic force microscopy (MFM) has high sensitivity and nanoscale resolution but the fields from the scanning tips can affect the intrinsic properties of the magnets [15, 16]. Diamond-based nano-magnetometry emerges as a promising approach to studying 2D magnets owing to its sub-nanotesla sensitivity, nanoscale resolution, and zero-field working condition [17-20]. However, there is a fundamental limitation in methods measuring the stray fields (such as SQUID, MFM and diamond magnetometry), that is, the stray fields vanish on the sample surface unless the detection location is close to edges or domain walls [21-23], so their measurement does not yield direct information about the intrinsic magnetization in the bulk of the materials but requires interpretation considering the edge and/or domain-wall effects.

In this work, we develop a method to measure the stray fields from arbitrarily designatable locations in thin-film magnets by integrating diamond-based nano-magnetometry with nano-indentation by atomic-force microscope (AFM) [24-26]. The key idea is to use nano-indentation to induce leakage of stray fields from the bulk of materials and then measure them using nanodiamond (ND) magnetometers [27]. We successfully detected the magnetic fields from designated locations in the bulk of both single-domain and multi-domain Fe thin films with nanoscale resolution and in a reversible and non-invasive manner. By comparing finite-element numerical simulation and experimental data, we extracted spatially resolved magnetization of the materials. This method can be applied to a broad range of materials, including atomically thin magnets [28-30] .

**RESULTS**

**Scheme for inducing and detecting stray fields from thin-film nanomagnets.** In a magnetic thin film, the magnetic field is confined within the material except for the leakage from edges or domain walls (as shown in Fig. 1a). We use the nano-tip of an AFM to induce nano-indentation of a magnetic flake so that stray fields can leak out at the indentation location from the bulk of the magnet (Fig. 1b). Then, the optically detected magnetic resonance (ODMR) of nitrogen-vacancy (NV) centers in an ND that is placed near the nano-indentation is employed for vector-magnetometry of the stray fields. We constructed a home-built AFM-ODMR correlated microscopy setup (Fig. 1c, for more details, see Methods and Supplementary Fig. S1) for the measurement.

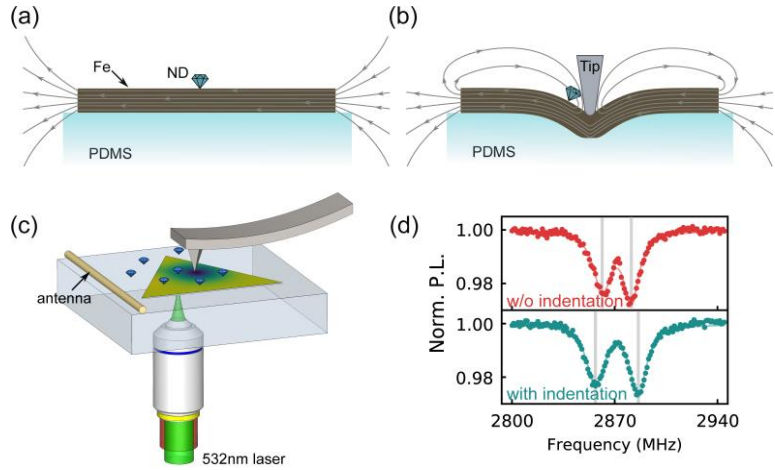

**Fig. 1. Scheme for inducing and measuring stray fields from the bulk of thin film materials.** (a) The stray field from a thin-film magnet vanishes near the surface at the bulk of the sample (where the nanodiamond sensor is placed). (b) An AFM tip is utilized to compress the thin magnet, inducing deformation that allows the magnetic field to escape the material. The leaked stray field can be measured by an ND magnetometer near the indentation location. (c) Schematic of the experimental setup. A confocal microscope sends laser to and collects fluorescence from an ND, and an antenna carries microwaves to control NV spins in the ND. A magnetic thin film is positioned atop a PDMS substrate layer. An AFM tip working in the contact mode applies nano-indentation on the magnet at arbitrarily designatable locations. (d) Example ODMR spectra of an ND on a Fe nanoflake under zero external field, recorded without (upper) and with (lower) nano-indentation. The indentation is induced by the AFM tip under a 300 nN force at 500 nm away from the ND.

We applied the method to Fe thin films. The Fe nanoflakes, synthesized through Te-assisted chemical vapor deposition (CVD) [31], were transferred to a polydimethylsiloxane (PDMS) substrate (see Methods for sample preparation). The NDs were uniformly dispersed on the material surface without agglomeration by spraying an ND solution (see Methods for ND parameters and preparation). The ND sensors can be moved to locations of interest on the sample by AFM tips working in the contact mode (see Supplementary Fig. S2). The energy level between the $|m_s = +1\rangle$ and $|m_s = -1\rangle$ states of the
5

NV spin is split by frequency $f \approx 2\gamma_e |B|\cos\phi$, where $\phi$ is the angle between the NV axis and the magnetic field $\boldsymbol{B}$, and $\gamma_e \approx 28$ MHz mT$^{-1}$ is the electron gyromagnetic ratio. By measuring the resonance frequencies of NV centers in ND with four different crystallographic orientations, both the magnitude and the direction (relative to the ND crystal) of the magnetic field $\boldsymbol{B}$ (which contains both the external field $\boldsymbol{B}_{\text{ext}}$ and the stray field from the material $\boldsymbol{B}_{\text{m}}$) can be determined. For reference, we first positioned an ND at the center of a Fe nanoflake under zero external field ($\boldsymbol{B}_{\text{ext}} = \boldsymbol{0}$) and recorded the ODMR spectrum of the ND without applying nano-indentation (upper panel of Fig. 1d). Subsequently, we measured the ODMR spectrum with nano-indentation applied near the ND (lower panel of Fig. 1d). Enhancement of the ODMR splitting due to the nano-indentation is clearly observed, which indicates the stray field leakage from the bulk sample. The process under the force is in the reversible elastic deformation ranges of both the Fe nanoflake and the PDMS substrate (see Supplementary Fig. S3) and the deformation length scale (about 80 nm) is much larger than the atomic size of the materials (with negligible effects on the microscopic interaction between spins in the material). This means the ND magnetometry of the nano-indentation induced stray field provides controllable access to the intrinsic magnetic properties of materials.

**The stray fields from a single-domain Fe thin film induced by nano-indentation**. We used a truncated triangular single-domain Fe nanoflake as a demonstration. Figs. 2a and 2b present the AFM and MFM images of the Fe thin film, respectively. The Fe nanoflake has a lateral size of about 8 μm and thickness approximately 15 nm, as characterized by the AFM imaging. NDs were scattered on the surface of the sample. We identified an ND with high ODMR contrast (indicated by the yellow arrow in Fig. 2a) and moved it close to the center of the Fe film using AFM nano-manipulation. Due to the strong strain in the NDs, the stray field from the indented Fe thin film was not strong enough to split the ODMR peaks of the NV centers along the four crystallographic orientations (see, e.g., Fig. 1d, where the ODMR spectra were measured without an external field applied), which makes vector-magnetometry insensitive under zero external field. To reduce the effects of the strain, we applied a magnetic field (about $52.8 \pm 0.3$ Gauss, calibrated by a reference ND outside the film, see Supplementary Fig. S4) perpendicular to the surface of the Fe thin film. We first measured the ODMR spectrum without applying the nano-indentation (Fig. 2c). The magnetic field $\boldsymbol{B}_1$ is obtained by fitting the resonance peaks (see Supplementary Note 1) with sensitivity about $0.056$ Gauss/Hz$^{1/2}$ (estimated in Methods). The field $\boldsymbol{B}_2$ under AFM indentation was obtained similarly. The magnitude difference $\Delta B = B_1 - B_2$ reflects the influence of indentation.

Since the stray field is significantly weaker than the external magnetic field (about 52.8 Gauss), the indentation primarily induces variations in the magnitude of the total magnetic field, with only minimal changes in its direction (less than 4° from the field without indentation in ND coordinate system, see Supplementary Fig. S5). Note that the change of the field direction relative to the ND crystal contains the rotation of ND caused by the deformation, which is estimated to be less than 2° (see the numerical simulation results in Supplementary Fig. S6). The measured field magnitude difference $\Delta B$ is mostly in the direction perpendicular to the sample, while the in-plane components of the stray field generated by the AFM indentation contribute only (negligible) secondary effects.

We measured the field change $\Delta B$ induced by nano-indentation at various designated locations around an ND sensor. The applied force was limited up to 300 nN so that the nano-indentation was reversible. To avoid potential artifacts induced by the tip-ND interaction, the AFM tip was positioned at least 500 nm away from the ND without any physical contact. We applied the nano-indentation at eight equally spaced directions (D1 to D8, indicated by colored circles in insets of Figs. 2d-2f) surrounding the ND. Figure 2d shows that $\Delta B$ oscillates with the direction of indentation locations. This oscillation can be understood by assuming that the magnetization is within the plane of the Fe thin film and the stray field at the ND sensor position is the strongest/weakest when the nano-indentation is in the directions (anti-)parallel/perpendicular to the magnetization. The field magnitude difference $\Delta B$ decreases with decreasing the indentation force or increasing the distance between the ND and the indentation location, as shown in Figs. 2e and 2f, respectively. Such dependence on the force and distance is consistent with the assumption that the field difference is due to the stray field leakage from the bulk of the Fe thin film induced by the indentation. The stray field due to the indentation vanishes at distances above one micron (see Fig. 2f), which means that if the sample is larger than a few microns the intrinsic fields from the bulk can be detected without the complications of edge effects.

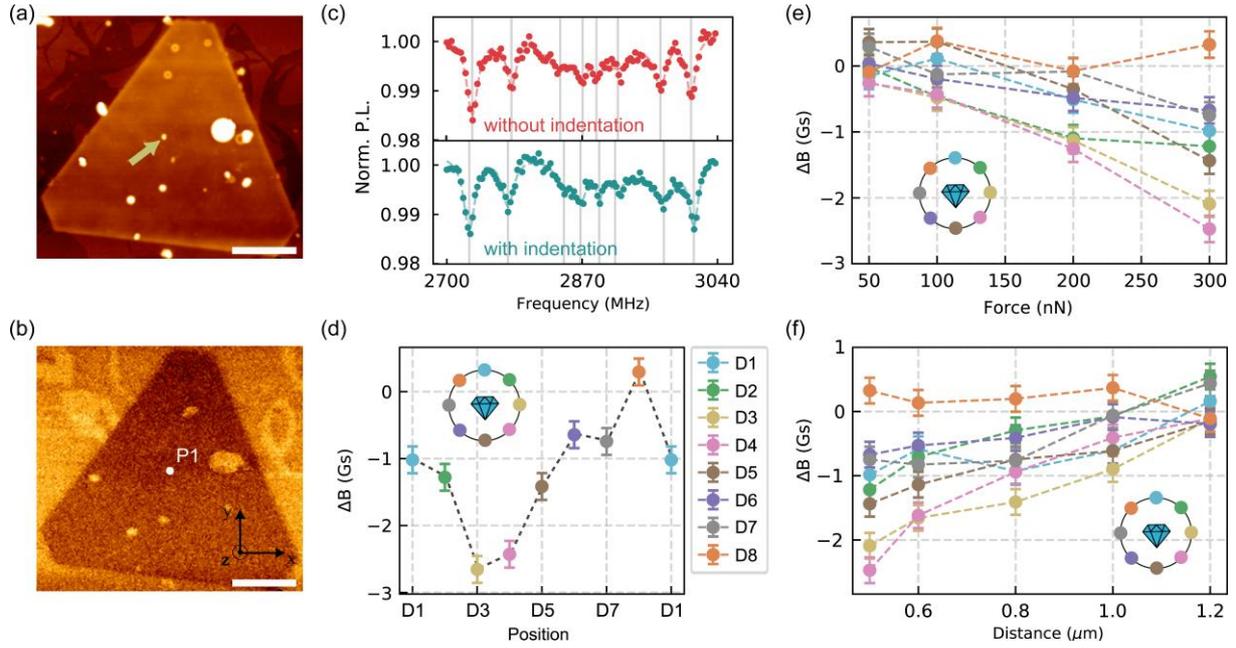

**Fig. 2. Nanodiamond-based sensing of stray fields from a single-domain Fe thin film induced by nano-indentation.** (a) AFM and (b) MFM images of a truncated triangular Fe thin film. The scale bars are 2 μm. The yellow arrow in (a) points to the ND sensor [marked by a circle symbol P1 in (b)] used as a nano-magnetometer. The inset in (b) shows the coordinate system adopted throughout the paper. (c) ODMR spectra of the ND, recorded without and with indentation, under a bias magnetic field of 52.8 Gauss applied perpendicular to the Fe thin film. The indentation is positioned 500 nm away from the ND, under a force of 300 nN. (d) The field difference $\Delta B$ measured by the ND magnetometer as a function of indentation direction (D1-D8, each increased by 45° clockwise from the $y$-axis, indicated by correspondingly colored circles in inset). All indentations were located 500 nm away from the ND, with an applied force of 300 nN. (e) $\Delta B$ as a function of indentation force applied at the same locations as in (d). (f) $\Delta B$ as a function of distance between the ND and the indentation locations along eight directions [marked in inset the same as in (d)], with the force being 300 nN. The error bars in (d-f) are fitting errors.

**Detecting stray field from a two-domain Fe nanoflake.** Figs. 3a and 3b present the AFM and MFM images of a Fe thin film with two domains, respectively. The thickness of this film is approximately 15 nm, as characterized by AFM imaging. An ND sensor with a size of approximately 100 nm was moved to different locations (subsequently from P1 to P7) on the Fe film. For each position of the ND, the field magnitude difference $\Delta B$ was measured with nano-indentation applied at eight directions at distance 500 nm from the ND with a force 300 nN (similar to experiments in Fig. 2d). For all positions of the ND, the field magnitude difference exhibits similar oscillations with varying the direction of the indentation. The oscillations have similar amplitudes but opposite signs when the ND is in the two different domains (Figs. 3c and 3d), which indicates that the magnetizations in the two domains have similar strength but opposite directions. The measurement was carried out subsequently for the ND sensor at locations from P1 to P7, during which the ND was moved across the domain walls three times (P1 to P2, P4 to P5, and P6 to P7). The agreement between different curves for the ND at different locations in the same domain suggests that indentation-induction of stray field is reversible and the domain structure is robust against the localized mechanical deformation. The fact that the field changes decrease with increasing distance between the ND and the indentation, as shown in Figs. 3e and 3f, is consistent with the assumption that the stray field is induced by the nano-indentation.

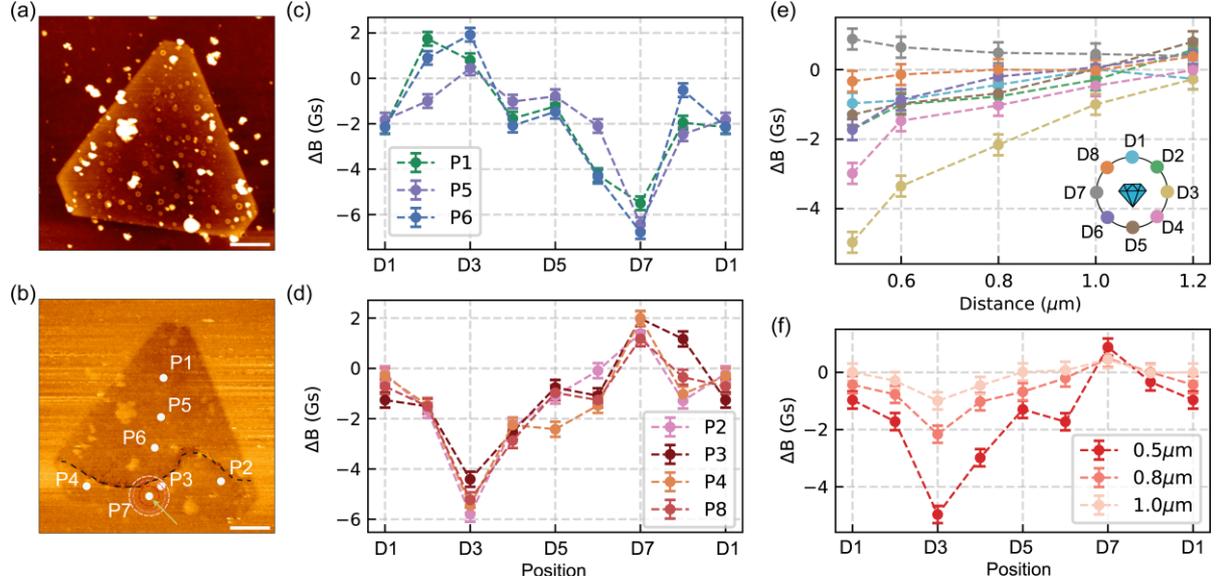

**Fig. 3. Detecting stray fields from different domains of a Fe thin film.** (a) AFM and (b) MFM images of the Fe thin film. The scale bars are 2 μm. The black dashed curve in (b) is a guide for the eye to the domain wall. (c, d) Field change $\Delta B$ as a function of indentation direction (D1-D8, defined relative to the ND, the same as in Fig. 2) when the ND was placed at different locations (P1, P5 and P6 in domain 1, P2, P3, P4 and P7 in domain 2). The indentation was applied at 500 nm away from the ND with a force of 300 nN. (e) $\Delta B$ as a function of distance of indentation from the ND (placed at P7) for different indentation directions (D1-D8, shown in inset, the same as in Fig. 2). (f) $\Delta B$ as a function of indentation direction [D1-D8, the same as in (d) and (e)] for various indentation distance [0.5 μm, 0.8 μm, and 1.0 μm, indicated by correspondingly colored circles abound P7 in (b)] from the ND (placed at P7). The error bars in (c-f) are fitting errors.

**Comparison between experimental data and numerical simulation.** We performed finite-element simulation of the deformation of Fe thin-films under nano-indentation and the stray field leakage. We compared the computed results with experimental data to extract information about the internal magnetization in the bulk of the materials. The deformation of the Fe thin film-PDMS bilayer system under AFM nano-indentation was computed with a model that incorporates rigorous contact mechanics at the nanoscale interface. The stray fields at the ND sensor position were obtained by summing the dipole fields from the finite elements of the deformed magnets, with the magnetization $\boldsymbol{M}$ assumed as a constant in each domain (see Methods for more details). Following the procedure used in actual experiments, we numerically simulated the deformation of the materials under indentation at various directions and distance around the ND and in turn the stray field at the position of the ND. The field magnitude $\Delta B$ was determined by the difference between the fields at the position of the ND with and without indentation applied.

In the simulation, the vector magnetization $\boldsymbol{M}$ was the only fitting variable for best reproducing the experimental data of $\Delta B$ as functions of indentations parameters (force, distance, and direction). As shown in Fig. 4a, the experimental data obtained from a single-domain Fe thin film are consistent with the numerical results, with the magnetization fitted to have magnitude $11{,}000 \pm 1{,}600$ A/m and an in-plane orientation of $298° \pm 9°$ clockwise from the $y$ axis. The data obtained from the two-domain sample are also consistent with the numerical simulation (as shown in Fig. 4b). The fitted magnetization has a magnitude of $18{,}000 \pm 3{,}600$ A/m at all locations, and the in-plane angle of the magnetization is fitted to be $72° \pm 12°$ at P1 and $283° \pm 11°$ at P3, nearly opposite to each other. The fitting results of magnetization at other locations (P2 and P4-P7) and more details are given in Supplementary Fig. S8.

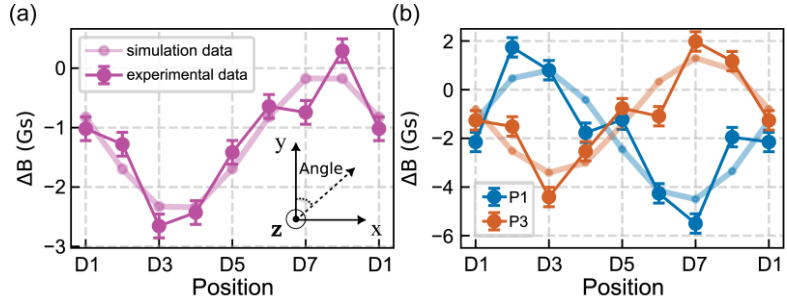

**Fig. 4. Comparison between experimentally measured field change and numerical simulation.** (a) Comparison for the single-domain sample (as in Fig. 2). In the simulation, the in-plane angle (clockwise from the $y$ axis shown in inset) of the magnetization in the flake is $298° \pm 9°$, with a magnetization intensity of approximately $11,000 \pm 1,600$ A/m. (b) Comparison for the two-domain sample (as in Fig. 3). The data and simulation are for P1 and P3. In the simulation, the in-plane angles [clockwise from the $y$ axis shown in inset of (a)] of the magnetization are $72° \pm 12°$ at P1 and $283° \pm 11°$ at P3 respectively, and the magnetization intensity is $18,000 \pm 3,600$ A/m. The error bars are fitting errors.

## CONCLUSIONS

We have demonstrated the feasibility of probing the intrinsic magnetization in the bulk of thin-film materials via indentation-induced leakage of the stray fields and nanodiamond-based magnetometry. The nano-indentation is reversible, and the deformation length has a size much larger than the atomic scale, which means that the microscopic mechanisms in thin-film magnets can be controllably studied without being complicated by external disturbance. On the other hand, the indentation has a spatial resolution of tens of nanometers, which can facilitate probing intrinsic properties in different regions of materials with high spatial resolution. The stray field leakage due to nano-indentation is concentrated near the indentation, so the intrinsic fields from the bulk can be extracted without complications due to the edge or domain wall effects. Indeed, we managed to measure the different bulk magnetization in two domains of a Fe nanoflake. The method developed here can be applied to atomically thin (2D) materials, where edge effects could fundamentally change the magnetic properties of the materials [32-37] and therefore approaches to directly probing the bulk magnetization are highly desirable.

## METHODS

**AFM-ODMR setup**

The experiments were conducted using an integrated platform consisting of an AFM and a confocal ODMR measurement system (see Supplementary Fig. S1 for a schematic). The ODMR measurement was carried out with a laser-scanning confocal system. NV center spins were excited by a 532 nm laser and controlled via microwaves transmitted through a copper wire antenna with a diameter of 20 μm. Spatial correlation between the AFM and confocal fluorescence measurements was established by overlapping the AFM image with the fluorescence image of NDs on the sample. The AFM imaging provides accurate coordinates for ND location and indentation locations, with spatial resolution determined by the tip radius (~ 30 nm).

For the indentation, an AFM (JPK Nano Wizard 4 XP BioScience) with an RTESPA-150-30 tip was employed all the time. For each indentation, a loading force was applied, and upon reaching the predetermined setpoint, a pulse signal was sent from the Pulse Generator (Pulse Streamer 8/2, Swabian Instruments) to initiate the cw ODMR measurements, which typically lasted 250-300 seconds. Throughout the entire indentation the probe height was continuously monitored in real time to ensure that no external disturbances interfered with the indentation. Detailed information regarding the variation in indentation force over the duration of the indentation and the AFM local depth-loading curves can be found in Supplementary Fig. S3.

**AFM calibration and MFM measurement**

We calibrated the AFM nano-indentation system through two independent procedures. (1) We determined the tip deflection sensitivity by performing indentation measurements on reference sapphire samples, and (2) we obtained the spring constant via thermal tuning of the cantilever. A MESP-V2 Si cantilever coated with a CoCr film, with a normal resonance frequency of 75 kHz and a spring constant of 2.8 N/m, was utilized for MFM imaging. During the MFM measurement, a lift height of 50 nm was consistently maintained.

**Preparation of PDMS substrates**

PDMS films were fabricated using the Sylgard 184 elastomer kit from Dow Corning. The silicone base and crosslinker were meticulously blended in a 10:1 weight ratio and degassed under vacuum. To prepare samples for AFM indentation experiments, the mixture was spin-coated onto a cover glass at 3000 RPM for 30 seconds and then cured at 60 °C for 24 hours. The film thickness typically measures around 50 μm, as determined by confocal microscopy, with the PDMS surface roughness below 6 nm as assessed via AFM imaging.

**Synthesis of Fe nanoflakes**

The Fe nanoflakes were synthesized on mica substrates using ambient-pressure CVD, following a procedure adapted from previous work [31]. The synthesis was carried out in a three-zone furnace (Lindberg/Blue M), with temperature independently controllable in each zone to ensure precise thermal regulation. Specifically, 20 mg of Te powder was positioned in the first heating zone, and 60 mg of $FeCl_2$ powder, loaded into a quartz boat, was placed in the second heating zone. Two freshly cleaved mica substrates were stacked together and positioned downstream of the $FeCl_2$ source. Then, the furnace chamber was purged with high-purity argon (Ar) gas at a flow rate of 300 sccm for 5 minutes to ensure an oxygen-free environment. Subsequently, a gas mixture of 200 sccm Ar and 20 sccm $H_2$ was introduced into the CVD system. Following these preparatory steps, the furnace was heated from room temperature to target temperatures within 25 minutes. The target temperatures were set as 530 °C in the first zone and 720 °C in the second and third zones. The growth was initiated after the target temperatures were reached in the three zones and lasted for 10 minutes. Then, the furnace was cooled down naturally to room temperature. Finally, the Fe nanoflakes were obtained between the two stacked mica sheets.

**Transfer of Fe nanoflakes to PDMS substrates**

The Fe nanoflakes were transferred to PDMS substrates by a polystyrene (PS)-assisted method. Specifically, 1.3 g of polystyrene was dissolved in 10 mL of toluene at 60 °C to form a homogeneous solution. A small amount of this solution was spin-coated onto the mica substrate containing the as-grown Fe nanoflakes. The mica was then heated at 130 °C to ensure proper adhesion of the PS film. Subsequently, the mica substrate with the PS-coated Fe nanoflakes was immersed in water, and the Fe nanoflakes, along with the PS film, were carefully detached from the mica using tweezers and a needle and then transferred onto a PDMS substrate. The assembly was heated at 130 °C to promote adhesion between the Fe nanoflakes and the PDMS. Finally, the PS film was dissolved by immersing the sample in toluene, leaving the Fe nanoflakes directly on the PDMS substrate.

**Preparation of ND sensors**

The ND solution was purchased from FND Biotech. A typical ND has a diameter of 100 nm and contains over 1000 NV centers. The ND solution was diluted in ethanol to a concentration of 1 µg mL$^{-1}$ and then ultrasonicated for 5 minutes to avoid potential aggregation. Subsequently, 200 µL of the ND dispersion was loaded into an atomizer (Paasche-HS202S) using a calibrated micropipette. The atomizer's pressure was optimized to 1 bar, generating a monodisperse aerosol that deposited the NDs uniformly across the sample surface. This gas-phase delivery method does not have the problem of ND aggregation resulting from drying in liquid-phase deposition.

After the deposition of NDs onto the sample surface, the cover glass bearing the sample was affixed to a PCB board. A microwave antenna was soldered onto the sample surface to connect with the microwave transmission lines on the PCB board.

**Sensitivity estimation**

The shot-noise-limited sensitivity of an NV magnetometer by cw-ODMR is [38]

$$\eta_{\text{cw}} = \frac{4}{3\sqrt{3}} \frac{1}{\gamma_e} \frac{\Delta \nu}{C_{\text{cw}} \sqrt{R}},$$

where $\gamma_e = 2.8 \text{ MHz/G}$ is the gyromagnetic ratio of the NV center. In our experiments, the photon-detection rate $R$ is about 2000,000 counts per second, the ODMR linewidth $\Delta \nu$ is 8 MHz, and the cw-ODMR contrast $C_{\text{cw}} = 0.028$. From these parameters, we estimate the magnetometry sensitivity to be 0.056 Gauss/$\sqrt{\text{Hz}}$.

**Finite-element simulation of deformation and stray fields**

The simulation was performed using COMSOL Multiphysics. The mesh was refined to 1–10 nm in regions including the iron film, the AFM tip-film interface, and the PDMS–film interface, while a coarser mesh (~100 nm) was applied elsewhere. The bottom surface of the PDMS was set as a fixed constraint. A force equivalent to the experimentally applied value was imposed on the top surface of the AFM tip.

The Fe film was assigned a Poisson's ratio $\gamma_1 = 0.29$ [39] with its thickness of 15 nm and Young's modulus $E_1 = 200$ GPa, determined via AFM topographic imaging and the nano-indentation measurements. The PDMS substrate (10:1 base-to-curing agent ratio) was characterized by AFM nano-indentation to have Young's modulus $E_2 = 2.8$ MPa. The Poisson's ratio of the PDMS substrate, $\gamma_2 = 0.4950$, was adopted from Ref. [40], where the same type of PDMS as in our study was measured. In the simulation, the indentation tip was assumed to have a conic shape with radius $R = 30$ nm and half-angle $\theta = 25°$, which follow the specifications of the Bruker tip RTESPA-150-30 used in the experiments. The AFM tip is made of antimony-doped silicon, with a Young's modulus of approximately 130 GPa and a Poisson's ratio of about 0.28 [41].

The numerically calculated deformation of the Fe thin film-PDMS bilayer system under AFM indentation is shown in the Supplementary Fig. S6 and Fig. S7, which reveal that the range of indentation is about 400 nm as determined by the distance of the maximum deformation curvature from the indentation center.

The spatial coordinates of the indented film were recorded with a resolution of 1 nm. Each discretized unit was assigned a magnetization vector *M*, treated as a point-like magnetic moment. The total magnetic field vector at the NV center position was then calculated by summing the contributions from all individual magnetic moments across the entire deformed film.

## ACKNOWLEDGEMENTS

This work was supported by Quantum Science and Technology-National Science and Technology Major Project no.2023ZD0300600, the Hong Kong Research Grants Council/Theme-based Research Scheme Project T45-406/23-R, the Hong Kong Research Grants Council Senior Research Fellow Scheme Project SRFS2223-4S01, the Research Grant Council of HKSAR Collaborative research fund (C4004-23GF) and the New Cornerstone Science Foundation. Y.H. and L.P. acknowledge the funding support from the National Key Research and Development Program of China (Nos. 2022YFA1203902, 2022YFA1204003).

**AUTHOR CONTRIBUTIONS**

R.B.L. and Q.L. conceived the idea and supervised the project, M.Z.A, K.Y.L., Q.L., and R.B.L. designed the experiments, M.Z.A constructed the setup with the contribution from W.H.L, Y.G., Y.C. and G.Z., M.Z.A and K.Y.L. performed the experiments, B.Z. synthesized the Fe thin films under supervision of Y.H. and L.P., M.Z.A, K.Y.L, Q.L., and R.B.L. analyzed the data, M.Z.A, K.Y.L., B.Z., Q.L., R.B.L. wrote the paper with the help of Y.H. and L.P, and all authors commented on the paper.

**REFFERENCES**